%% file: paper1.tex
\documentclass[pre,twocolumn,english]{revtex4}
\usepackage[T1]{fontenc}
\usepackage[latin9]{inputenc}
\usepackage{amsthm}
\usepackage{amsmath}
\usepackage{graphicx}

\makeatletter
\@ifundefined{textcolor}{}
{%
 \definecolor{BLACK}{gray}{0}
 \definecolor{WHITE}{gray}{1}
 \definecolor{RED}{rgb}{1,0,0}
 \definecolor{GREEN}{rgb}{0,1,0}
 \definecolor{BLUE}{rgb}{0,0,1}
 \definecolor{CYAN}{cmyk}{1,0,0,0}
 \definecolor{MAGENTA}{cmyk}{0,1,0,0}
 \definecolor{YELLOW}{cmyk}{0,0,1,0}
 }

\usepackage{amsmath}
\usepackage{bm}
\usepackage{graphicx}
\usepackage{color}
\usepackage{epsfig}
\usepackage{epstopdf}

\makeatother

\usepackage{babel}

\begin{document}

\title{Monte-Carlo study of scaling exponents of\\
rough surfaces and correlated percolation}

\author{I. Mandre}

\affiliation{Institute of Cybernetics at Tallinn University of Technology, Akadeemia
tee 21, 12618, Tallinn, Estonia}

\author{J. Kalda}

\affiliation{Institute of Cybernetics at Tallinn University of Technology, Akadeemia
tee 21, 12618, Tallinn, Estonia}
\begin{abstract}
We calculate the scaling exponents of the two-dimensional correlated
percolation cluster's hull and unscreened perimeter. Correlations
are introduced through an underlying correlated random potential,
which is used to define the state of bonds of a two-dimensional bond percolation model. 
Monte-Carlo simulations are run and the values of the scaling exponents are determined
as functions of the Hurst exponent $H$ in the range $-0.75\leq H\leq1$.
The results confirm the conjectures of earlier studies.
\end{abstract}
\maketitle

\section{Introduction}

\begin{figure}
\noindent \begin{centering}
\includegraphics[width=0.9\columnwidth]{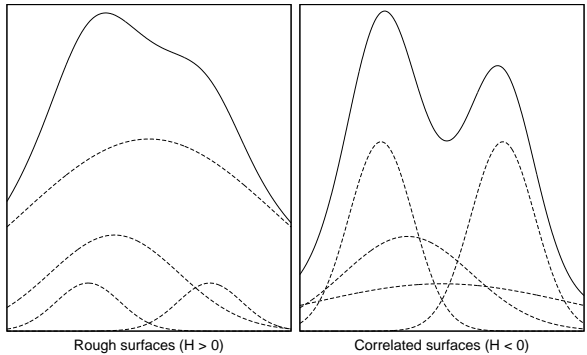}
\par\end{centering}

\caption{\label{fig:H-interpretation}The potential \eqref{eq:multiscale-as-sum}
is made up of components with different amplitudes. The amplitude
of a component of scale $\lambda$ is proportional to $\lambda^{H}$.
For $H>0$ the wider {}``hills'' start to dominate the landscape
and once $H\geq1$ only the widest {}``hill'' matters. Conversely,
for  $H<0$ local fluctuations start to gain in influence and once
$H<-0.75$ the wider hills lose any influence on the scaling exponents
of the surace.}

\end{figure}

\begin{figure}
\noindent \begin{centering}
\includegraphics[width=0.9\columnwidth]{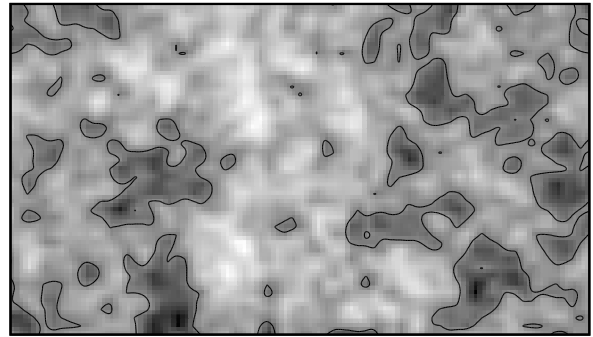}
\par\end{centering}

\caption{\label{fig:example-potential-with-isolines}An example of a random
potential ($H=0$ and so $D_h=1.5$ \citep{Kondev1,Kondev2}) with isolines separating the area into {}``land'' and
{}``sea''.}

\end{figure}

The world around us is chaotic and seemingly
random, but one can still find regularities. Take mountain ranges
--- these irregular jagged structures may seem intractable to analysis,
but often display an interesting property --- they look the same at
different length scales --- they are self-similar. This phenomenon
is not limited to the surface features of planets, but is also found in many
other places --- deposited metal films \citep{Palasantzas1994}, ripple-wave
turbulence \citep{Wright1997}, crack fronts in material science \citep{Bouchaud,Santucci2007,Bakke2008},
cloud perimeters \citep{Pelletier1997}, passive tracers in two-dimensional
fluid flows \citep{Catrakis1,Kondev3,Isichenko1991,Bernard2007}, etc.

The analysis of these physical systems is often reduced to determining
the scaling exponents characterizing the rough surfaces involved.
For example, the convective-diffusive transport of a passive scalar
in a random two-dimensional steady flow is determined by the scaling
exponent of the isolines of the underlying stream-function \citep{Isichenko1991}.

The aim of this paper is to numerically calculate the values
of these exponents depending on the roughness parameter $H$.
In Sec.~\ref{sec:overview}, we start off by giving an overview of the concepts used --- rough
surfaces and percolation clusters, what we mean by correlations,
and a mapping between the two classes of models.
Numeric calculations are done through the Monte-Carlo simulations
using the two-dimensional bond percolation model; the procedure is
described in Sec.~\ref{sec:monte-carlo}. Interpretation of
the resulting data requires overcoming the finite size effects and
the convergence problems, which is the subject of Sec.~\ref{sec:data-analysis}.
Finally, Sec.~\ref{sec:conclusion} provides a brief summary of the results and
a future outlook for the studies of the statistical topography of random surfaces.

\section{\label{sec:overview}Overview}

\textbf{Rough surfaces.}
Let $\psi(x,y)\equiv\psi(\mathbf{r})$ be the height or potential
of a self-similar random two-dimensional surface. We define the roughness
exponent $H$ --- also known as the Hurst exponent --- through the
surface height drop at distance $a=\left|\mathbf{a}\right|$ \citep{Mandelbrot}:\begin{equation}
\left\langle \left[\psi(\mathbf{r})-\psi(\mathbf{r}+\mathbf{a})\right]^{2}\right\rangle \propto\left|\mathbf{a}\right|^{2H},\label{eq:surface-height-drop-law}\end{equation}
where angular braces denote averaging over the coordinate $\mathbf{r}$
(or also over an ensemble of surfaces). This scaling law assumes that
$a_{0}\leq\left|\mathbf{a}\right|\leq a_{1}$, where $a_0$ and $a_1$ are the lower and upper cut-off scales, and $0\leq H\leq1$.
Relation \eqref{eq:surface-height-drop-law} also describes self-similarity
--- the height of a {}``hill'' on the surface is a power law of
its diameter, so hills at different scales have the same proportions.

A more generic description, which is not limited to the positive values
of $H$, can be given through the power spectrum $P_{\mathbf{k}}$
\citep{MBI}. We assume that\begin{equation}
\left\langle \psi_{\mathbf{k}}\right\rangle =0,\quad\left\langle \psi_{\mathbf{k}}\psi_{\mathbf{k}'}\right\rangle =P_{\mathbf{k}}\delta_{\mathbf{k}+\mathbf{k}'},\end{equation}
and define the spectral density $P_{\mathbf{k}}$ as a power law:\begin{equation}
P_{\mathbf{k}}\propto\left|\mathbf{k}\right|^{-2H-2},\quad\mbox{for }\left|\mathbf{k}\right|\ll a_{0}^{-1}.\label{eq:power-spectrum-def}\end{equation}
This also allows us to conveniently divide the {}``multiscale''
potential $\psi(\mathbf{r})$ into a sum of {}``monoscale'' functions,\begin{equation}
\psi(\mathbf{r})=\sum_{\lambda_{i}=2^{i}\lambda_{0}}\psi_{\lambda_{i}}(\mathbf{r}),\label{eq:multiscale-as-sum}\end{equation}
where each component in the sum represents a function with a single
characteristic length. The effect of the parameter $H$ on the amplitude
of the ``monoscale'' components of the ``multiscale'' potential can be seen in Fig.~\ref{fig:H-interpretation}.

The potential $\psi$ can be also characterized through a correlation
function (covariance) \begin{eqnarray}
C(\mathbf{a}) & = & \left\langle \psi(\mathbf{x})\right\rangle \left\langle \psi(\mathbf{x}+\mathbf{a})\right\rangle .\label{eq:potential-covariance}\end{eqnarray}
Indeed, for potentials conforming to \eqref{eq:power-spectrum-def}, it is
a power law\begin{equation}
C(\mathbf{a})\propto|\mathbf{a}|^{2H}\qquad a_{0}\ll|\mathbf{a}|\ll a_{1},\label{eq:potential-covariance-power-law}\end{equation}
which is valid for the range $-3/4\leq H\leq1$.

With a specific height $h$, the random potential defines a set of isolines $\psi(\mathbf{r})=h$
(see Fig.~\ref{fig:example-potential-with-isolines} for an example).
The height $h$ can be interpreted as the ``sea level''. So, a
single isoline can be looked at as the coastline of an island or a
lake. The coastline is also a self-similar structure, and one can
quickly run into difficulties when trying to measure its length ---
the result depends on the size of the measuring stick used \citep{Mandelbrot1967}.
The parameter that best characterizes isolines is their scaling exponent
--- also called their fractal dimension $D_{h}$ --- which is a fractional
number. The length of a coastline can then be given through\begin{equation}
\left\langle L(\lambda)\right\rangle \propto\lambda^{1-D_{h}},\end{equation}
where $\lambda$ is the size of the measuring stick. $D_{h}$ is a
nontrivial function of the underlying surface's Hurst exponent:\begin{equation}
D_{h}=D_{h}(H).\end{equation}
Alternatively, we can take two points at a distance $a$ on a coastline,
and calculate the length of the line between them, assuming a fixed
measuring stick $\lambda=\lambda_{0}$:\begin{equation}
\left\langle L(a)\right\rangle \propto a^{D_{h}}.\label{eq:scaling-law}\end{equation}

\begin{figure}
\noindent \begin{centering}
\includegraphics[width=0.45\columnwidth]{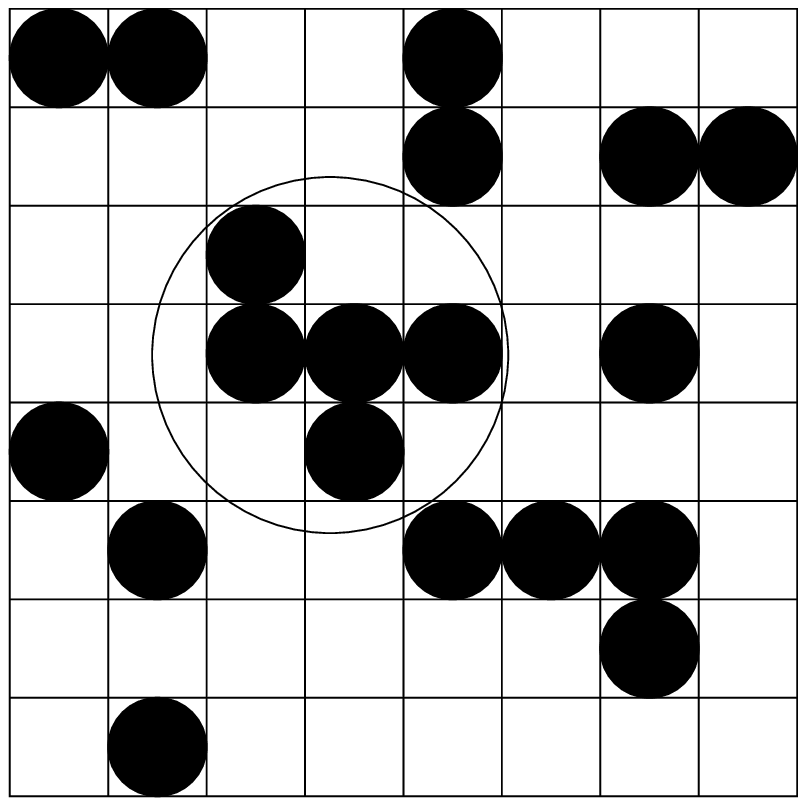}\includegraphics[width=0.45\columnwidth]{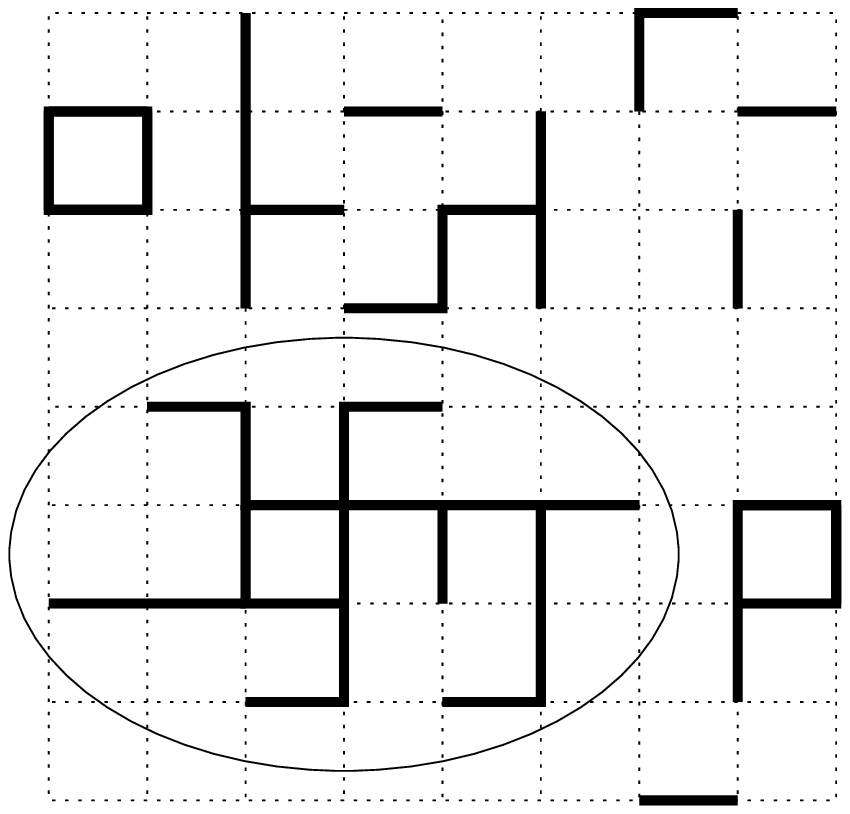}
\par\end{centering}

\caption{\label{fig:percolation-examples}Examples of regular percolation lattices.
Square lattice site model to the left and square lattice bond model
to the right. Largest clusters have been circled.}

\end{figure}

\textbf{The percolation problem} is concerned with the structures
that form by randomly placing elementary geometrical objects (spheres,
sticks, sites, bonds, etc.) either freely into continuum, or into
a fixed lattice (Fig.~\ref{fig:percolation-examples}). Two objects
are said to communicate, if their distance is less than some given
$\lambda_{0}$, and communicating objects form bigger structures called
clusters. Percolation theory studies the formation of clusters and
their properties. The more interesting aspect is when and how is
an infinite cluster formed. This depends on the lattice site occupation
probability $\eta$. The minimum site occupation probability when
an infinite cluster appears is called the percolation threshold $\eta_{c}$.
Near this probability, the percolation model displays a critical behavior
and long-range correlations.

Percolation theory is used to study and model a wide variety of phenomena:
from a fluid flow in a porous medium to thermal phase transitions
and critical behavior in magnetism with dilute Ising models.

\begin{figure}
\noindent \begin{centering}
\includegraphics[width=0.5\columnwidth]{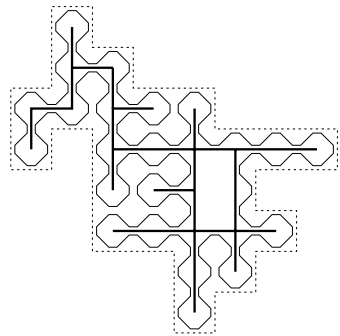}
\par\end{centering}

\caption{\label{fig:one-cluster}An example square bond percolation cluster.
The cluster is made up of the bold segments, the zig-zag is the hull
and the dashed line is the unscreened perimeter.}

\end{figure}

The structures that can be identified in conjunction with a percolation
cluster are the cluster itself, the hull and the unscreened perimeter
(Fig.~\ref{fig:one-cluster}). Aside from these one more structure
can be identified --- the oceanic coastline \citep{JK08EPL}. It is
comprised of the hull of a cluster and also the hulls of all the holes
(reverse clusters) within it. Near the percolation threshold, all of
these structures (the cluster, the hull, the unscreened perimeter
and the oceanic coastline) are fractals and can be characterized by
scaling exponents. 

Looking at a percolation cluster and an isoline of a random potential,
we can identify similar structures: one can look at the percolation
cluster as an island and its hull as the coastline of the island.
This idea will be fleshed out in more precise terms farther below,
where a mapping between the two models is described.

One very important and useful aspect about the scaling exponents is
a phenomenon known as universality \citep{MBI} --- within specific
universality classes, the scaling exponents take the same value across
different percolation models. More specifically, they are invariant
to small fluctuations or distortions in the lattice structure (for
instance, decaying exponentially to distance). This means that the
scaling exponents for both the random square bond lattice, and the
random square site lattice are the same: the both models belong to
the same universality class of the two-dimensional uncorrelated percolation.

\textbf{Correlated percolation.} Percolation lattice does not need
to be completely random, but can entail certain correlations. Here
we describe the percolation lattice through an infinite set of random
variables $\theta_{i}$ which are unity at occupied sites and zero
at empty sites ($i$ denotes the site number). Then we can characterize
the correlations through the correlation function\begin{equation}
c_{\theta}(\mathbf{x}_{i}-\mathbf{x}_{j})=\left\langle \theta_{i}\theta_{j}\right\rangle -p^{2},\label{eq:correlation-function-sitevalue}\end{equation}
where $p=\left\langle \theta_{i}\right\rangle $ is the site occupation
probability.

Alternatively \citep{Weinrib2}, correlations can be brought into
the percolation model by assigning each lattice site a random number
$p_{i}\in[0,1]$ where $\left\langle p_{i}\right\rangle =p$. The
site values are then calculated as\begin{equation}
\theta_{i}=\Theta(p_{i}-x_{i}),\label{eq:weinrib-generation-forumla}\end{equation}
where $\Theta(x)$ is the Heaviside step function and $\left\{ x_{i}\right\} $
are independent random variables uniformly distributed in $[0,1]$.
The correlation function is\begin{equation}
c_{p}(\mathbf{x}_{i}-\mathbf{x}_{j})=\left\langle p_{i}p_{j}\right\rangle -p^{2}\end{equation}
and $c_{\theta}(\mathbf{a})=c_{p}(\mathbf{a})=c(\mathbf{a})$ \citep{MBI}.

We are interested in algebraically decaying correlations so that\begin{equation}
c(\mathbf{a})\propto\left|\mathbf{a}\right|^{2H},\quad H\leq0.\label{eq:correlated-perc-power-law}\end{equation}
It is believed \citep{MBI} that the universality class is determined
by the two-point correlation function. One can show that at $H<-3/4$,
the model belongs to the universality class of uncorrelated percolation
\citep{Weinrib2}. However, in the range $-3/4\leq H\leq0$, the correlations
do affect the scaling exponents.

\begin{figure}
\noindent \begin{centering}
\includegraphics[width=0.95\columnwidth]{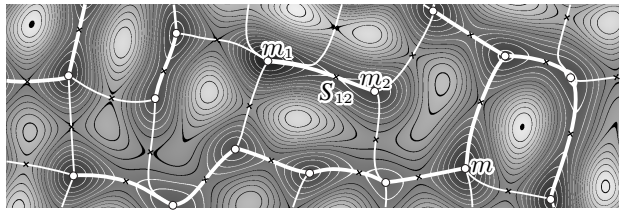}
\par\end{centering}

\caption{\label{fig:mapping}Mapping between the random surfaces and the percolation
problem. The bond between two maxima $m_{1}$ and $m_{2}$ is connecting,
if the saddle point $S_{12}$ on them is above the flood level. So,
saddle points are mapped into percolation bonds.}

\end{figure}

\textbf{There exists a simple mapping between the rough surfaces and
the percolation model} \citep{Ziman,Weinrib1}. According to it, the local
maxima of the potential define the lattice sites and the lattice bonds
are obtained by drawing fastest-ascent paths from all the saddle points
(Fig.~\ref{fig:mapping}). The surface $\psi(\mathbf{x})$ is {}``flooded''
at a given level $h$ and a bond $i$ is left connecting if the saddle
point $\mathbf{x}_{i}$ on it is above the water (is land), that is
when $\psi(\mathbf{x}_{i})\geq h$. As a result, we get an irregular
two-dimensional lattice; recall that as per universality, small distortions
of the lattice don't affect the resulting scaling exponents.
With this mapping, we can relate the islands
formed at flooding to the resulting clusters, and their coastlines
to the hulls of the said clusters.  Also,
	if the surface correlation function \eqref{eq:potential-covariance}
is a power law \eqref{eq:potential-covariance-power-law}, then so
is the correlation function \eqref{eq:correlation-function-sitevalue}
for the percolation model \eqref{eq:correlated-perc-power-law}, where
the parameter $H$ is the same. Due to universality, the scaling exponents
of the matching structures are also the same.

\begin{figure}
\noindent \begin{centering}
\includegraphics[width=0.95\columnwidth]{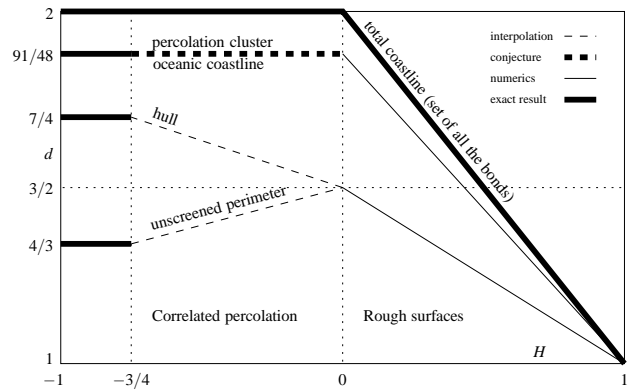}
\par\end{centering}

\caption{\label{fig:previous-results}Known results, conjectures and interpolations
for the scaling exponents as functions of the Hurst exponent $H$.
While these functions can be approximated as linear functions, they
are in fact non-trivial.}

\end{figure}

The scaling exponents are of interest in many applications so,
there is a need to calculate their values depending on the underlying
surface's roughness parameter $H$. In Fig.~\ref{fig:previous-results},
we can see the known results (numeric and analytic), and also interpolations
and conjectures for the range $-3/4\leq H\leq0$. Our next task is
to run simulations to numerically shed light on these gray areas (the
scaling exponents of the hull and the unscreened perimeter). As these
exponents behave the same way for the both problems of rough surfaces and 
correlated percolation, we can calculate
them using the model which is the most convenient from the numerical point of view.

\section{\label{sec:monte-carlo}Monte-Carlo simulations}

\textbf{Generation of percolation clusters.} We calculate the scaling
exponents using the model of correlated two-dimensional  bond
percolation on square lattices. Our first task is to generate said percolation models
so that they conform to the correlation function \eqref{eq:correlated-perc-power-law}.
For this, we first generate random potentials with the requested roughness
$H$. We take the ``flood level'' as the value of the potential
at some starting location $\mathbf{x}_{0}$, so that $h=\psi(\mathbf{x}_{0})$,
and use it to map the random surface model into one of the percolation
models. This is done by overlaying the percolation lattice on the
rough surface and calculating the bond values using the Heaviside
step function as\begin{equation}
\theta_{i}=\Theta\left(\psi(\mathbf{x}_{i})-\psi(\mathbf{x}_{0})\right).\end{equation}
This approach is similar to the one described by equation \eqref{eq:weinrib-generation-forumla}
and preserves the correlation exponent. It is slightly different from
the maxima-saddle point mapping, but due to universality, there will
be no change in the values of the scaling exponents. Using the potential
as the underlying model allows us to get the results in the parameter
range $-0.75\leq H\leq1$.

\begin{figure}
\noindent \begin{centering}
\includegraphics[width=0.8\columnwidth]{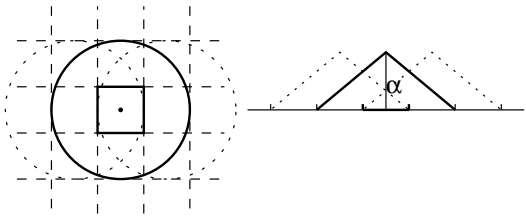}
\par\end{centering}

\caption{\label{fig:conef}For the $i$-th layer $\psi_{\lambda_{i}}(\mathbf{r})$
in \eqref{eq:multiscale-as-sum}, we divide space into a grid of cells
with side length of $\lambda_{i}=2^{i}$ ($\lambda_{0}=1$). At the
center of each cell $j$, a cone is placed with height $\alpha_{j}=r_{j}2^{iH}$,
where $r_{j}$ is an independent uniform random variable in the range
$[-0.5,0.5]$. The cones have a diameter of $3\lambda_{i}$ and so
overlap.}

\end{figure}

\textbf{To generate random potentials,} we exploit formula \eqref{eq:multiscale-as-sum}.
We generate different components (layers) of the potential for different
lengths and sum them up. The $i$-th layer $\psi_{\lambda_{i}}(\mathbf{r})$
is formed by a grid of cells, where the grid cell side length is $2^{i}$,
and at the center of each cell $j$ we place a cone of height $r_{j}2^{iH}$,
where $r_{j}$ is an independent uniform random variable in the range
$[-0.5,0.5]$ (Fig.~\ref{fig:conef}). The cones have a diameter
of $3\lambda_{i}$, so the resulting overlap yields a smoother potential.
However, the overlap or the shape used (cone, in this case) does not
significantly affect how fast the results converge to the asymptotic power laws. So,
one can choose a shape more optimized for the speed of numerical calculations (for instance,
one could use a simple block or a cylinder). The number of layers
required depends on the value of $H$. For example, at $H=0$, we
can work with $i=0\ldots 30$.

\begin{figure}
\noindent \begin{centering}
\includegraphics[width=0.8\columnwidth]{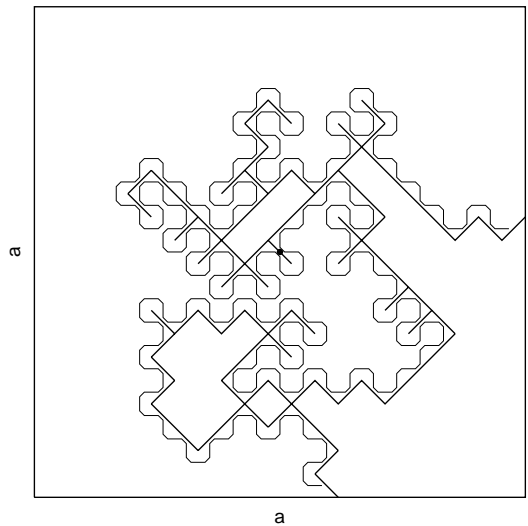}
\par\end{centering}

\caption{\label{fig:hull-tracing}Calculating the length of a hull for size
$a$. We start from the center of a $a\times a$ box and dynamically
calculate bond values as we trace the hull until reaching the side
of the box. We discard hulls that make a full circle. The box sizes
picked are $a=8,16,32,\ldots,1024,\ldots$.}

\end{figure}

\textbf{Gathering data} is a matter of generating percolation clusters
of different sizes and tracing the structures of interest within these.
For the hull, this is done in the following way. We constrain ourselves
to a $a\times a$ box, and start to dynamically trace the hull from
the center point $\mathbf{x}_{0}$ until it reaches any sides of the
box (Fig.~\ref{fig:hull-tracing}). Hulls that make a full circle
are discarded. To get additional data, we also trace backwards from
the center, so that the both ends of the hull reach the box's sides. We
do this millions of times for differently sized boxes $a=8,16,\ldots,1024,\ldots$.
While for some cases, the range of values $a=8,16,\ldots,512$ is sufficient,
slow convergence in some regions of $H$ forces us to use larger lattices.
However, the size is limited by computational resources and in our
case it was not practical to go over $a=2048$.

\begin{figure}
\noindent \centering{}\includegraphics[width=0.8\columnwidth]{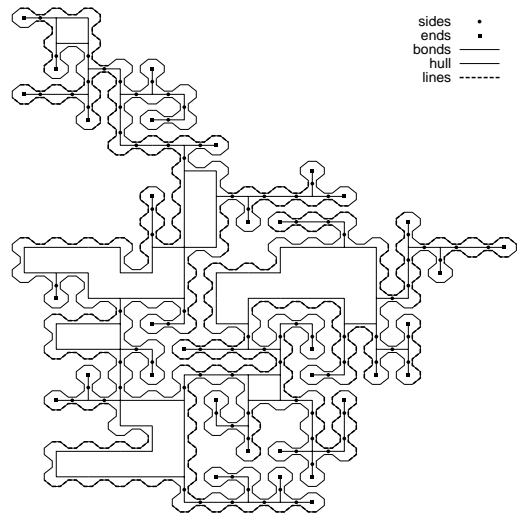}\caption{\label{fig:hull-parts}Counting the size of the hull can be done in
many ways. One can simply count the number of segments (hull), the
number of bonds the hull touches from both sides (sides), the total
number of bounds the hull touches (bonds), the number of times a single
bond sticks out (ends), or the number of times the hull forms a straight
line of 4 segments (lines). All of these scale with the same power
law.}

\end{figure}

\begin{figure}
\noindent \begin{centering}
\includegraphics[width=0.8\columnwidth]{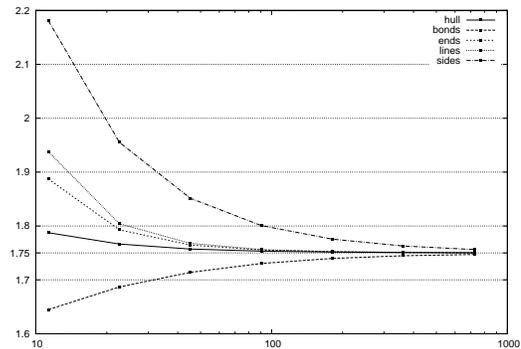}
\par\end{centering}

\caption{\label{fig:uncorr-conv}Convergence of the hull properties for the
uncorrelated percolation (towards $D_{h}=7/4=1.75$).}

\end{figure}

Once we have a hull, we want to measure its size (length). While the
easiest way would be to just sum the number of segments in the trace
line, it is also possible to determine the size by other properties
(Fig.~\ref{fig:hull-parts}). All of these properties scale with
the same power law. We can denote the size of a single hull as $L_{i}(a_{j})$,
where $i$ indicates the property. As this is different for each individual
hull, we find the average value \begin{equation}
\mathcal{L}_{i}(a_{j})=\left\langle L_{i}(a_{j})\right\rangle ,\end{equation}
and as per equation \eqref{eq:scaling-law}, this should scale as
a power law with the exponent $D_{h}$. However, the scaling is asymptotic
($a\rightarrow\infty$), so for finite $a$, there are sizable deviations
called finite size effects. These come from the geometry and finite
size of the lattice. Using \eqref{eq:scaling-law}, we can estimate
the exponent as\begin{equation}
\tilde{D}_{h}\left(\sqrt{a_{j}a_{j-1}}\right)\simeq\ln_{2}\frac{\mathcal{L}(a_{j})}{\mathcal{L}(a_{j-1})}\quad(a_{j}=2a_{j-1}).\end{equation}
Plotting this for the uncorrelated percolation (Fig.~\ref{fig:uncorr-conv}),
we can see how the different properties converge towards the value
$D_{h}=7/4$. The finite size effects are strongly manifested for
the smaller lattices.

\section{\label{sec:data-analysis}Data analysis}

\textbf{To calculate the scaling exponents} from the data, the following
assumptions are made:
\begin{enumerate}
\item the mathematical expectation for each property can be described as
an infinite series\begin{equation}
\overline{L}_{i}(a)=\sum_{\mu=1}^{\infty}A_{i\mu}a^{\alpha_{i\mu}},\qquad\alpha_{i(\mu+1)}<\alpha_{i\mu},\label{eq:property-as-infinite-series}\end{equation}
where $i=1,\ldots,m$;
\item $\alpha_{i\mu}\equiv\alpha_{\mu}$ for $\mu=1,\ldots,m$;
\item the leading terms in the sum are linearly independent ($\det\left\Vert A_{\mu i}\right\Vert \neq0$).
\end{enumerate}
After these assumptions, we apply a variation of the least squares
method described in \citep{finsizesc} and previously used in \citep{JK08EPL}.
The method works if the assumptions made are correct (the method also
validates them) and yields us the value $\alpha_{1}$, which is the
scaling exponent we are looking for. So, the reason why we counted
all the different properties for the hull (Fig.~\ref{fig:hull-parts}),
is that they are necessary for this method.

For the unscreened perimeter, a similar approach is taken. The unscreened
perimeter is obtained by taking a hull but pruning it from {}``fjords''.

\begin{figure}
\noindent \begin{centering}
\includegraphics[width=0.8\columnwidth]{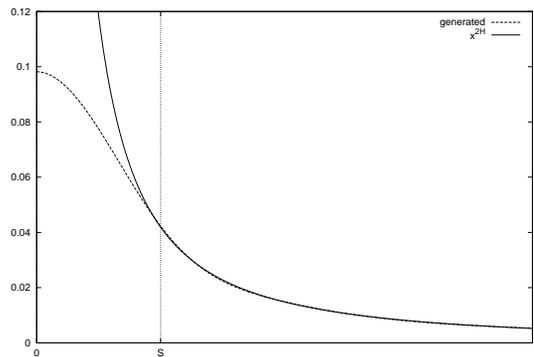}
\par\end{centering}

\caption{\label{fig:covariance-compared-to-powerlaw}Covariance of a potential
versus the power law $x^{2H}$. Here $S$ marks the length at which
the covariance starts to diverge from the power law.}

\end{figure}

\begin{figure}
\noindent \begin{centering}
\includegraphics[width=0.8\columnwidth]{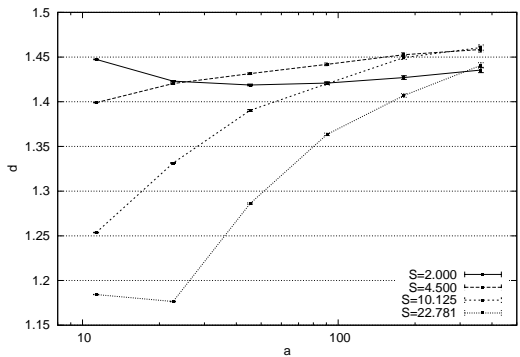}
\par\end{centering}

\caption{\label{fig:convergence}Convergence of a property of the unscreened
perimeter at $H=0$ (towards theoretically known $D_{u}=1.5$) for
different values of parameter $S$.}

\end{figure}

\textbf{Convergence problems.} In some areas the calculations are
hindered by very slow convergence. Let parameter $S$ represent the
length at which the covariance of the generated potential starts to
differ from that of the ideal $|\mathbf{a}|^{2H}$ law:\begin{equation}
S=3\cdot2^{s-1},\end{equation}
where $s$ indicates the smallest-scale layer index (Fig.~\ref{fig:covariance-compared-to-powerlaw}).
When $S$ decreases (by adding bottom layers), local fluctuations
start to gain in influence compared to those of the long-range correlations.
This causes a strong finite-size effect and the scaling exponents
behave as if $H$ was smaller (Fig.~\ref{fig:convergence}). Conversely,
when increasing $S$, the scaling exponents can initially behave as
if $H$ was greater than it really is.

To get over these distortions, one could calculate for bigger lattice
sizes. But often this is not an option as convergence can be very
slow and computational resources are limited. Another way would be
to manually find the optimal layer configuration that minimizes distortions.
This is the approach we took and yielded good results for the hull
(aside from the values $H=-0.75$ and $H=+1.00$). However, the convergence
of the unscreened perimeter is very sensitive to changes in the layer
configuration and for most data points did not yield clear results.

\textbf{The results} can be seen in Fig.~\ref{fig:results-plot}
and Tab.~\ref{tab:results-table}. The hull behaves as expected.
While it did not yield clear results at $H=-0.75$ and $H=1.00$,
the extrapolations provided seem to indicate that it terminates at
$7/4$ and $1$ respectively. Due to the convergence problems, the results for the unscreened perimeter
are not as clear. However, one can say that at least that the result
do not contradict the analytical findings and support the applicability of a nearly-linear interpolation between the points
$D_u(0)=1.5$ and $D_u(-3/4)=4/3$.

\section{\label{sec:conclusion}Conclusion}

We have run Monte-Carlo simulations to determine the scaling exponents
of the hull and the unscreened perimeter as functions of the Hurst
exponent in the range $-0.75\leq H\leq1$. 
For this, we first generated
random potentials conforming to the required correlation function
by summation of component potentials of different characteristic lengths
and mapping the potential into percolation models. Hulls and unscreened
perimeters were traced from these models, and their lengths calculated
for different scales. A variation of the least squares method was
used to obtain the values of the exponents. 

\begin{figure*}
\noindent \begin{centering}
\includegraphics[width=0.95\textwidth]{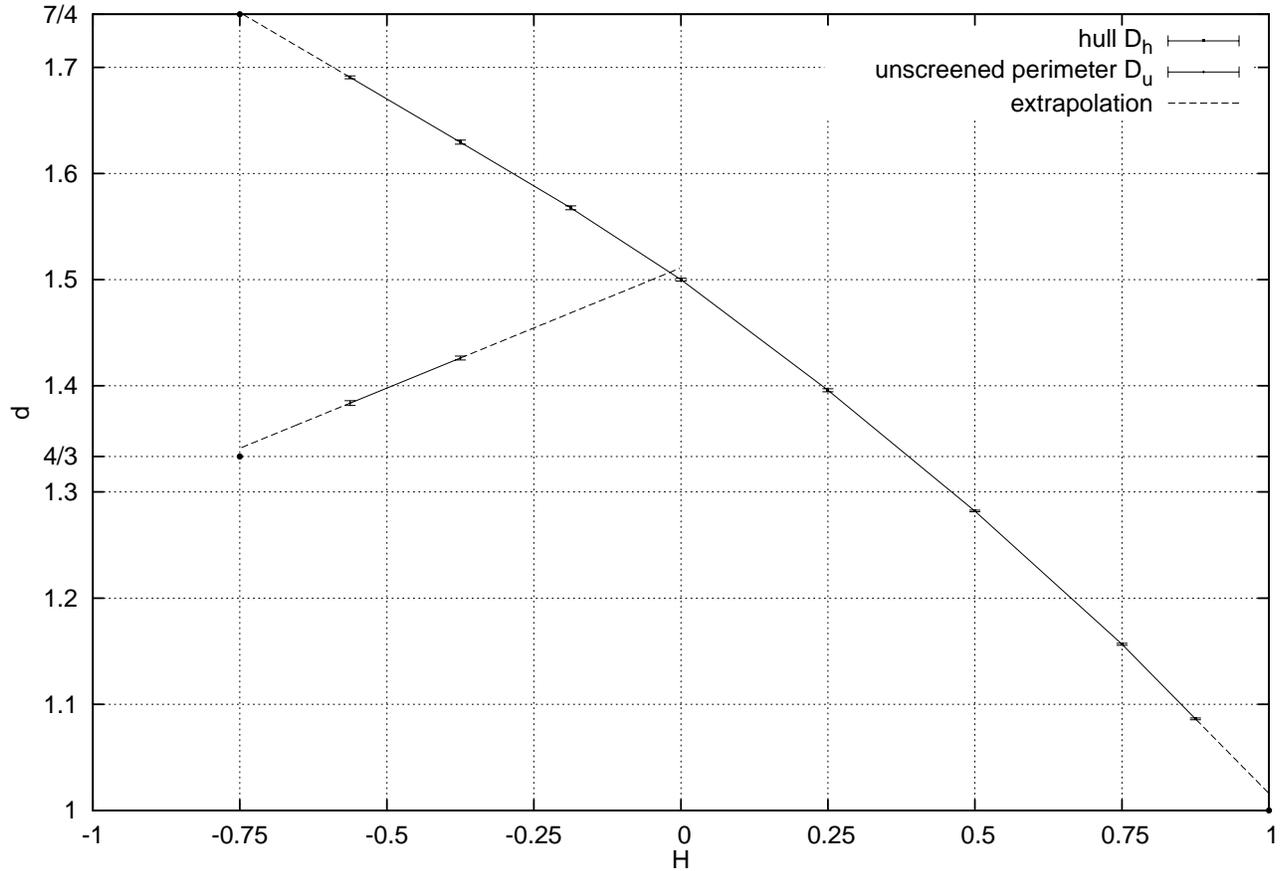}
\par\end{centering}

\caption{\label{fig:results-plot}Scaling exponents of the hull and the unscreened
perimeter as functions of the Hurst exponent $H$. Data points for
the positive side of $D_{u}$ are not plotted as to avoid clutter.}

\end{figure*}

The results confirm the
previously known data in the range $0\leq H\leq 1$ and also the conjectures
for the behavior in the range $-0.75\leq H\leq0$, see Fig.~\ref{fig:results-plot}.
The particular results regarding the fractal dimension $D_h(H)$ of hulls 
confirm that for $0\leq H\leq 1$, the 4-vertex model (i.e. the rough surfaces in 1+1-dimensional geometry)  \citep{4VM}
belongs to the same universality class as the isotropic Gaussian self-affine surfaces (assuming the respective equality of the Hurst exponents).
Indeed, comparing the numerical results of Ref.~ \citep{4VM} and those of the current study shows that 
for the entire range of $0\leq H\leq 1$, the values of $D_h(H)$ coincide within the uncertainties of {\em ca} $10^{-3}$.
An  important consequence is that the conjecture about the super-universality of the 
loop correlation exponent $x_l(H)\equiv\frac 12$ \citep{Kondev1,Kondev2} (which has been exploited in several studies, c.f. \citep{Rajabpour2009,Saberi2010,Nezhadhaghighi2011,Ramisetti2011})
is clearly rejected: this conjecture implies $D_h(H)=\frac 32 -\frac H2$, which falls well beyond the uncertainty 
margins of the present simulation results  (for instance,  at $H=\frac 12$, the conjectured value $\frac 54$ falls far 
from the range of $1.2820\pm 0.0008$).

The obtained results are valuable for a range of practical applications (such as the turbulent transport
in quasi-stationary velocity fields), for which the scaling exponents have been analytically expressed via the fractal
dimension of the hull. As a future outlook, our method can be applied to calculate other scaling exponents of the correlated percolation
problem and statistical topography, such as the fractal dimensions of the clusters (oceanic coastlines), percolation backbone, etc. 
It can be also extended to study the scaling laws of the transport on quasi-stationary velocity fields.

\begin{table*}
\noindent \begin{centering}
\input{results_table.tex}
\par\end{centering}

\caption{\label{tab:results-table}Numeric results for the hull and the unscreened
perimeter ($0.95$ confidence).}

\end{table*}
\section{Acknowledgments}
This work was supported by Estonian Science Targeted Project No.
SF0140077s08 and Estonian Science Foundation Grant No. 7909.

\bibliographystyle{apsrev}
\addcontentsline{toc}{section}{\refname}\bibliography{paper1}

\end{document}

%% file: results_table.tex
\begin{tabular}{ |r|c|c| }
\hline
H\ \  & $D_h$ & $D_u$ \\ 
\hline
$1.0000$ & & \\
\hline
$0.8750$ & $1.0862 \pm 0.0008$ & $1.0862 \pm 0.0022$\\
\hline
$0.7500$ & $1.1565 \pm 0.0010$ & $1.1565 \pm 0.0011$\\
\hline
$0.5000$ & $1.2820 \pm 0.0008$ & $1.2820 \pm 0.0011$\\
\hline
$0.2500$ & $1.3958 \pm 0.0014$ & \\
\hline
$0.0000$ & $1.5000 \pm 0.0013$ & \\
\hline
$-0.1875$ & $1.5676 \pm 0.0018$ & \\
\hline
$-0.3750$ & $1.6295 \pm 0.0019$ & $1.4261 \pm 0.0018$\\
\hline
$-0.5625$ & $1.6906 \pm 0.0014$ & $1.3837 \pm 0.0023$\\
\hline
$-0.7500$ & & \\
\hline
\end{tabular}